\documentclass[12pt]{article}
\usepackage[]{epsfig}
\usepackage{epstopdf}
\newcommand{\be}{\begin{equation}}
\newcommand{\ee}{\end{equation}}
\newcommand{\bq}{\begin{eqnarray}}
\newcommand{\eq}{\end{eqnarray}}

\newcommand{\bc}{\begin{center}}
\newcommand{\ec}{\end{center}}

\def\ov{\overline}

\def\(({\left(}
\def\)){\right)}
\def\[[{\left[}
\def\]]{\right]}
\def\bi{\bibitem}

\begin{document}

\title{Replica method and finite volume corrections}
\author{Matteo Campellone, Giorgio Parisi, Miguel Angel Virasoro \\
Dipartimento di Fisica,  INFN,\\
Statistical Mechanics and Complexity Center (SMC) - INFM
  - CNR\\
Universit\`a di Roma ``La Sapienza''\\
Piazzale A. Moro 2, 00185 Rome (Italy)
}

\maketitle

\begin{abstract}
In this note we introduce a method to calculate the finite volume corrections to the mean field 
results for the free energy when  replica 
symmetry is broken  at one-step.  We find that the naive results are modified by the presence of
additional corrections: these corrections can be interpreted as arising from fluctuations in the size of the blocks in the replica approach.
The computation suggests a new approach for deriving the replica broken results in a rigorous way.

\end{abstract}


\section{Introduction}

The exact solution  of the mean field theory of a wide class of spin glasses and other disordered models can be 
found using the replica approach. 
When we lower the temperature these models undergo a phase transition:  
the system  freezes in a highly correlated phase without necessarily breaking any 
symmetry of the Hamiltonian (sometimes the Hamiltonian has no symmetry at all).  
The replica formalism allows us to describe this phase transition in a spontaneous symmetry breaking 
framework \cite{mpv,PABOOK}.  One introduces the replica symmetry that is spontaneously broken in the low temperature phase.

Although many of these results can be obtained in a rigorous mathematical way,  the original replica method is not crystal clear from a pure mathematical viewpoint.  We shall also see that there are ambiguities when we compute the finite volume corrections.
The analytic framework of the replica method needs a deeper comprehension  and this work should be a contribution in that direction.

There is a perfectly well understood physical interpretation of the phenomenon of replica symmetry 
breaking using a probabilistic approach: all the computations can be done in a transparent way 
without ever introducing a non-integer number of replicas \cite{mpv,PABOOK}.  
Unfortunately at the present moment this probabilistic approach does not allow us to calculate in a 
simple way the finite volume corrections.  In principle the replica method allows such a computation, 
but the naive results turned out to be wrong, if compared with the exact results for 
the random energy model (REM) of Derrida.  The absence in the conventional replica approach of a term 
corresponding to the fluctuations of the parameter $m$ was clearly identified as the likely source of 
this discrepancy \cite{theo,pafera}.

In this note we show that the results of the usual one step replica symmetry breaking solution can be obtained 
within a conventional saddle point approach if a given infinite class of saddle points is taken into 
consideration.  In this way we have a well defined mathematical setting where we can compute the 
finite volume corrections to the mean field results and derive results that are correct in the 
case of the REM. 

The usual parameter $m$ of replica symmetry breaking appears as an integration variable.  In the infinite volume limit  the $m$
 integral is rapidly oscillating and it is dominated by a saddle point in the variable 
$m$, where the integration path in the complex plane is perpendicular to the real axis.  This explains why the saddle point in the conventional replica approach is a minimum and not a maximum as function of 
$m$.  Of course we still have to justify why the sum over the particular class of saddle points we 
have chosen should give the exact results, however this is a well defined mathematical problem, that 
we will not address here.

\section{A short introduction to the replica method}

In disordered systems we are interested in computing the average free energy 
${F(\beta)}$ defined as
\be
{F(\beta)}= - \lim_{N \to \infty} F_N(\beta)\ , \ \ \
 F_N(\beta)\equiv ( N \beta)^{-1}\ov{ \ln \left( Z_N (\beta)\right) }\ ,
\ee
where the bar denotes the average of the random instances of the problem and $Z_N(\beta)$ is the partition function for a system with $N$ degrees of freedom at inverse temperature $\beta$.

The quantity ${F(\beta)}$ can also written as:
\be
 \lim_{n\to 0} F^{(n)}(\beta) \ ,
\ee
where
\be
F^{(n)}(\beta) = - \lim_{N \to \infty} (n N \beta)^{-1} \ln \left( \ov{Z_N(\beta)^n}\right)\,
\ee

In the replica formalism one introduces a temperature-dependent effective free energy
$F(Q_{ab})$, where $Q_{ab}$ is a symmetric matrix zero on the diagonal, the pairs of indexes $a,b$
assume  $n(n-1)/2$ values; eventually $n$ has to go to zero in the
calculation of the physical quantities. 
In many interesting cases one can derive the exact representation
\be
 \ov{Z_N(\beta)^n}=
 C(N,n) \int dQ \exp( -N \beta F(Q) ) \ ,
\ee
where $C(N,n) =(2 \pi N)^{n(n-1)/4}$. For lightening the notation we have not indicated the obvious dependence of $F$ on $n$ and $\beta$. Moreover in some models (depending on minor details \footnote{These corrections are absent in the Sherrington Kirkpatrick model for spin glasses, if the coupling are Gaussian random variables, but they are present in the case of bimodal distribution of the couplings.}), one should add subleading terms in $N$ in the exponent: we will not consider in this note this complication that is irrelevant for our aims (however it is crucial if one wants to obtain expressions correct at the subleading level in a given model).

The permutation group of $n$ elements acts naturally on the matrix $Q$, i.e. $(Q^\pi)_{a,b}=Q_{\pi(a),\pi(b)}$.
The function $F(Q)$, that depends on the model, is invariant under the action of the permutation group (the so called replica group).

If we use formally the method of the point of maximum (that is {\em justified} in the limit $N \to \infty$), the  free energy density is given by
\be
\ov{F}=\lim_{n\to 0}F(Q^{*})\ ,\label{EQ}
\ee
where $Q^{*}$ is the (supposed unique) stable  solution of the equation
\be
{\partial F(Q) \over \partial Q_{a,b}}=0 \ .
\ee
More precisely a solution  of the previous is stable if the Hessian matrix 
\be
{\cal H}_{a,b;c,d}\equiv{\partial F(Q) \over \partial Q_{a,b}  \partial Q_{c,d}}
\ee
 is a non-negative matrix.

If we solve the previous equations for integer $n$, the solution is an $n \times n$ matrix (no 
analytic continuation is needed at this stage).  
For integer $n$ the maximum    solution is always given by a matrix
where all off-diagonal elements are equal, { i.e.} $Q_{a \neq b}=q$; in this case the replica 
symmetry is exact (i.e. the solution of eq. (\ref{EQ}) is invariant under the permutations of $n$ elements): this is the replica symmetric ($RS$) solution.  The properties of this solution can be analytically continued in $n$ up to $n=0$, that is the interesting point. At low  
temperatures it may happen that the $RS$ solution is no longer stable for small $n$ ($n<1$).  This means that in the low temperature phase the relevant solutions eq. (\ref{EQ}) are not invariant under the replica group.  The way in which 
this replica-symmetry is broken depends on the particular model but basically two main universality 
classes have been identified in mean field models.

For a first class of models the mean field solution $Q$ has the one-step of replica symmetry-broken 
($1RSB$) form.  In this case the possible  values of $Q_{ab}$ are only two: 
$Q_{ab}=q_0$ or $Q_{ab}=q_1$ (with $q_{1} > q_{0}$).  In many of the models  $q_{1}\ne 0$  $q_{0}=0$.   For simplicity in this first approach to the problem of calculating finite size corrections we will only consider here the case where $q_{0}=0$.

For integer $n$ the $1RSB$ solution can be represented in terms of an $n \times 
n$ matrix with $n/m$ blocks of size $m \times m$ on the diagonal.  Outside the blocks $Q_{ab}=q_0$, 
while within the blocks $Q_{ab}=q_1$:
\bq
Q_{ab}=q_1, \ \mbox{if} \ \ \mbox{Int}(a/m)=\mbox{Int}(b/m) \ , \nonumber \\
Q_{ab}=q_0, \ \mbox{if} \ \ \mbox{Int}(a/m)\ne \mbox{Int}(b/m)\ . \label{ONESTEP}
\eq
 Eventually in the replica method also $m$ takes non-integer 
values. Models, where the replica symmetry is broken at the one step level, are interesting for their
relevance to the behaviour of structural glasses \cite{KWT,PVAR}.

For a second class of systems the low temperature solution can be parameterized in terms of a 
continuous function $q(x)$.  These models have a different phenomenology from models of the first 
class and the physical interpretation of the solution is more involved \cite{mpv,PABOOK,BOOK}.

In this note we will only 
consider models of the first class ($1RSB$) and we try to obtain a deeper comprehension of some points of the 
replica method in this simpler case.

A difficulty present in all models with a replica symmetry breaking transition ({\em i.e.} 
models of the first and of the second class) is  related to the analytic continuation in $n$. Indeed we have to  find the solution of eq.(\ref{EQ}) for $n=0$ and we  have to provide an Ansatz on the form of the matrix $Q$.  Of course $0 \times 0$ 
matrices do not exist in reality, but they may be considered as the analytic continuation up to 
$n=0$ of some matrices that are defined for convenient integer positive values of $n$ (one 
continues analytically not the matrix, but scalar functions of the matrices).  This space is huge 
and there is no definite number of parameters over which we can maximize the free energy to obtain the 
saddle point solution. A way to bypass this difficulty is to decide {\em a priori} the form of the solution with 
a fixed number of parameters: we restrict ourself to a particular 
family of solutions and determine the best solution within that family.

In models where the $1RSB$ solution is correct we assume that the solution to the saddle point 
equations is of the previous described form (eq.(\ref{ONESTEP})).  Under this assumption it is possible to compute the 
effective free energy as function of $q_0$, $q_1$, $n$ and $m$.  The form of the effective free energy can be 
analytically continued to $n=0$ and to generic $m$.  In this way one obtain an effective free energy 
$F(q_0,q_1,m)$ where $m$ is a real parameter (eventually the parameter $m$ turns out to belong to 
the interval $[0-1]$).  The solution to the saddle equation  can 
be found by extremizing the free energy with respect to $q_0$, $q_1$ and $m$.  In the three cases 
($q_{0}=q_{1}$, $m=0$ and $m=1$) we recover  the replica symmetric solution. It is remarkable that the 
free energy in the replica broken case is higher that the free energy in the replica symmetric case.

We acknowledge that in this approach we 
choose {\em a priori} the form of the solution and, unless we find an alternative way to solve exactly the 
particular model, there is no way to assure ourselves that there is no other solution, maybe 
completely different, that gives the exact free energy density.  

\section{A simple exact representation}

Let us consider a system with infinite range interactions where the mean field approach gives the 
correct results in the thermodynamic limit. As starting point
we follow backwards Derrida's approach to REM \cite{de} and use an integral representation of the logarithm to
calculate  the average free energy:

\be
\ov{\ln{Z_N}} = \int_{0}^{\infty}  {dt\over t} \left( \exp(-t)
- \ov{\exp(-t Z_N})\right) \ .
\label{reprln}
\ee
Let us define
\be
\exp(-\phi(t,N))\equiv\ov{\exp(-t Z_N})  \ .
\ee
We now perform a Taylor expansion around 0:
\be
\ov{\exp(-t Z_N})  = \sum_{k=0, \infty}{1 \over k!}(-t)^{k} \ov{Z_N^{k}}.\label{MIRA}
\ee
and compute $\ov{Z_N^{k}}$ using the representation mentioned in the previous section
\be
\ov{Z_N^{k}} =C(N,k)\int dQ \exp (-N \beta F(k,Q)) \ , \label{inte}
\ee
where the integral is done over the parameters of the $k \times k$ matrices.

The reader should notice that the Taylor expansion in eq. (\ref{MIRA}) is probably non-convergent: in the case of Gaussian disorder $\ov{Z_N^{k}}$ diverges
as  $\exp(A k^2)$ for large $k$. However we shall see later that this might not be a problem.

In this approach everything is written in terms of the average of the partition function to an integer 
power and therefore no analytic continuation is needed.  The value of $\ov{Z^{k}_N}$ could be evaluated in 
the large $N$ limit using a saddle point approximation in eq. (\ref{inte}).  However, when the 
volume $N$ goes to infinity, we have to evaluate the sum for large values of $t$ 
($\ln(t)\simeq N)$ and different terms may cancel.  Therefore we are not allowed to restrict ourselves to the 
leading estimate of the $\ov{Z^{k}_N}$. Nevertheless, if a sufficiently 
accurate evaluation of the the quantities $\ov{Z^{k}_N}$ is done for large $N$, we should obtain the 
correct result. We will conjecture that there is an effective way to do this computation.

\section{New conjectures}

As a consequence our task is to obtain the best approximation to the quantities $\ov{Z^{k}_N}$ in the large 
$N$ limit. We must obtain uniform approximations because the expansion in powers of $t$ in eq. (\ref{inte})
cannot be exchanged with the integral over $t$. The reader should notice that here we stick to integer $k$ and no analytic continuation in $k$ is done. Naively one could think that if we know exactly the function $\ov{Z^{k}_N}$ for all $k$, we also know its analytic continuation at non-integer $k$. However this is not evident in this particular case. Indeed if $A$ is a positive quantity, the knowledge of the moments 
\be
A^{(k)}\equiv \int_0^\infty d\mu(A) A^k
\ee
determines the positive measure $\mu(A)$ in a unique way only in some cases. In particular if \be
\sum_{k=1,\infty} \left(A^{(k)}\right)^{-1/k}=\infty\, ,\ee
 the measure is unique. 
Unfortunately in our case the $J$ have a Gaussian distribution \footnote{If the distribution of the coupling $J$ is bounded, the analytic continuation would be uniquely defined.} and the moments increase as $\exp( C k^2)$. The question of uniqueness is therefore open.

For each given $k$ there may be many stationary points of the argument of the exponent in equation 
(\ref{inte}) and the leading 
contribution when $N$ goes to infinity can be easily evaluated.  However, as already remarked, this 
is not sufficient because of the strong cancellations and subdominant terms must be taken into account.  
We find it convenient to make two conjectures that allows us to make further progress.

\begin{itemize}
\item We  conjecture  that if we approximate $\ov{Z^{k}_N}$ by the sum over {\sl all} 
the saddle points, this approximation is enough to obtain the correct results if inserted into eq.  
(\ref{MIRA}).  It may be not so simple to classify all the saddle points for the function 
$F(k,Q)$, although in some cases it is possible. We have the task of finding all the solutions for integer $k$ of the equation $\partial F(k,Q)/ \partial Q=0$.

\item
We further conjecture that in the case of one step replica symmetry breaking, the correct results are obtained if we only consider some saddle 
points that generalize the one step replica symmetry breaking.  
\end{itemize}
Let us be more specific. We will restrict our search to 
those matrices $Q$ that can be divided into blocks of size $m_i$, where $\sum_{i=1}^l m_i =k$ and $m_i>0$.  
(Here $l$ is the total number of blocks of the matrix.)  The off diagonal elements of $Q_{ab}$ have 
a constant value (that may be $l$ dependent) if $a$ and $b$ belong to the same block.  In other words 
\be
Q_{ab}=q_i \  \mbox{if} \ a\in B_i, \ b\in B_i \ .
\ee
Moreover we consider, 
as candidates for the stationary points, only matrices where $Q_{ab}$ is zero if $a$ and $b$ do not 
belong to the same block. This last requirement is reasonable 
in the case where it turns out that $q_{0}=0$ in the usual replica approach.

In this way each stationary point is characterized (apart from permutations) by the size of the blocks $B_i\equiv m_i$ and by the values of $q_i$.
The same contribution appears more than once. To determine its multiplicity we imagine performing all possible permutations of the k rows/columns and checking whether the matrix generated is different. For instance, if all $m_i$ are different, the contribution is:
\be
\frac{k!}{\prod_{i=1}^l m_i!}
\exp\left(- N \beta F(\{m\}) \right)
\delta(\sum_{i=1}^l m_i -k),
\ee
where $l$ is the number of blocks in the matrix $Q$ \footnote{For simplicity we have assumed that we have only one non-zero solution for the $\{q\}$ at fixed $\{m\}$.}.

\section{The leading term}
Let be more definite and let us put these conjectures at work. The prototype   models  we have in mind are spin glasses models with with a $p$-spin interaction.  For $p>2$ their low 
energy phase is described by one step replica symmetry breaking and in the limit $p\to \infty$ they 
coincide with a soluble model: Derrida's REM \cite{de,GM}. 

In these models one can verify through an explicit computation that at the saddle point  the leading term when $N$ goes to infinity is factorized into  contributions from each block.  Neglecting terms of order 1 the final 
expression for the contribution of a given saddle point is given by
\be
\exp\left(- N \beta F(\{m\}) \right)= \exp\left(-N \sum_{i=1,l} \beta m_i (f(m_i)) \right)  \ .
\ee

We must now sum over all the possible saddle points, counted with their multiplicity. The final
result is
\begin{equation}
\overline{Z_N^k} = \sum_{\nu_1,\nu_2...=0}^\infty \frac{k!}{\prod_{m=1}^\infty {\nu_m! (m!)^{\nu_m}}}\exp\left(-N\beta\sum_{m=1,\infty}{ \nu_{m} m f(m)}\right) \delta(\sum_m m \nu_m-k) ,
\label{zk}
\end{equation}
where $\nu_m$ is the number of blocks of size $m$.
The quantity
$-f(m)$ is equal to the $1RSB$ free energy whose form depends on the model. 
In the REM one verify that that one obtains the correct formulae using 
\be
f(x) = -\frac{\beta}{4} x- \frac{\ln(2)}{\beta x}\ .   \label{FREM}
\ee
Indeed in the particular case of the REM  the previous representation is exact  without subleading correction. The computation was done in an explicit way in \cite{de}.

We can now use the previous expressions for $\overline{Z_N^k}$ in the computation of $\exp(-\phi(t))$.
Using (\ref{zk}) we can write
\be
\exp(-\phi(t,N)) \equiv \sum_{k=0}^\infty \frac{(-t)^k}{k!} \overline{Z_N^k} =
\exp \left( \sum_{r=1}^\infty \frac{(-t)^r}{r!} \exp( - r N  \beta f(r)) \right).
\label{sum}
\ee

A detailed computation shows that the limit, where $N$ and $t$ both go to $ \infty$ at constant $y=\ln{t}/N$, is relevant to compute the 
average free energy (this is justified a-posteriori).  We just face the problem of evaluating in this region the quantity
 \be
 -\phi(t,N)=\sum_{r=1}^\infty \frac{(-t)^r}{r!} \exp( -r N \beta  f(r))\ .\label{PHI} \ .
 \ee

For this purpose we follow the method (introduced in this context by \cite{ca,MK}) of transforming 
the previous sum into an integral in the complex plane around the integers 
$r=0,1,2,3...\infty$ and then deform the contour of integration to obtain an integral in one 
variable that can be evaluated by the saddle point method in the complex plane.

In this way we obtain 
\be
-\phi(t,N) =   \frac{1}{2i} \int_C
\frac{ \exp(N (x y  - x \beta f(x)))}
{\Gamma[1+x] \sin[\pi x] } dx\, ,
\ee
where $y=\ln{t}/N$ and $C$ is an appropriate integration path in the complex plane: $C$ goes from $+\infty+i \epsilon$ to $+\infty -i\epsilon$ crossing the real line at $0<x<1$.  This path may be deformed by breaking it into smaller circles running counterclockwise around the positive integers so as to obtain the previous formula. We now deform it so that it goes from $-i\infty$ to $+i \infty$. The possibility of doing this deformation is not clear. However quite often the sum in eq. (\ref{PHI}) is not convergent and the rotation of the path in the complex plane may be a possible way of  giving a meaning to this non-convergent sum.

We now try to see what happens when $N$ goes to infinity and we separate the leading from the subleading terms. We look for a saddle point in the complex plane.  The equation for the saddle point (i.e. $x_{sp}$) is 
\be
\beta f(x_{sp}(y)) + x_{sp}(y) \beta f^\prime(x_{sp}(y)) - y = 0.
\label{mspa}
\ee
 Let us assume, for simplicity, that the leading contribution come from the region where  $0<x_{sp}<1$ \footnote{This is usually true at low temperatures, at high temperatures $x_{sp}>1$ and we stay in the unbroken replica phase.}; indeed in the saddle point approximation for the $t$ integral the dominating values of $y$ will be such that $0<x_{sp}(y^*)<1$. If the  In this case we have at the leading and first sub-leading order

\be
-\phi(t,N) = C(y) \exp \left(N x_{sp}(y) ( y - \beta f(x_{sp}(y))\right) \ ,
 \label{ephi}
\ee
and
\be
C(y) = \sqrt{1 \over 2 \pi N\beta}\frac{\Gamma[-x_{sp}(y)] \ }
{\sqrt{-2 f^{\prime}
(x_{sp}(y)) -  x_{sp}(y) f^{\prime \prime} (x_{sp}(y)) }} 
 \label{ephi1}
\ee
is a (positive) quantity  whose value is irrelevant to leading order in $N$, but will be useful in the next section.

In order to extract the leading order contribution we notice that for large $N$ 
\bq
\phi(t,N) \approx 0 \ \  \Longrightarrow \ \  \exp(-\phi(t,N))=1 \ \ \mbox{if} \ \ y -\beta f(x_{sp}(y))<0\ , \\ 
\phi(t,N) \approx \infty \ \  \Longrightarrow \ \  \exp(-\phi(t,N))=0 \ \ \mbox{if} \ \ y- \beta f(x_{sp}(y))>0\ .
\eq
In other words we can approximate $\exp(-\phi(t,N))$ with 0 or 1 depending on the sign of $y -\beta f(x_{sp}(y))$. This approximation being not valid in a region of with $1/N$ (in $y$) around the point 
\be
y^*= \beta f(x_{sp}(y))\ .
\label{eqlnt}
\ee
Finally we obtain

\be
\ov{\ln{Z_N}} = \int_{0}^{\infty}  {dt\over t} \left( \exp(-t)
- \exp(-\phi(t,N)\right) = N y^* +O(1)\ .
\label{fin}
\ee

Note that equations (\ref{mspa}) and  (\ref{eqlnt}) are equivalent to 
\be
f^\prime(x_{sp})=0\ ,
\ee
that is  the saddle point equation for the size of
the block of the $1RSB$ solution. In the replica approach the previous equation is derived
maximizing with respect to $x$ the function
$f(x)$ that has the
meaning, from equation (\ref{zk}),
of free energy density (per replica) of $x$ replicas in a state with overlap $q=q_1$. We have rederived  the usual equation of the replica approach with one step symmetry breaking following a different route.

We notice that the condition $x_{sp}<1$ is equivalent to the condition $T<T_c$. It is possible
that fluctuations over $x$ are connected to
sample-to-sample fluctuations of the critical temperature. In any case for  $x_{sp} > 1$
{\em i.e.} $T>T_c$, the form of the finite-$N$ corrections  is
different because they are due to the existence of other saddle
points that have a weight that is proportional to
 $\exp{- N \mu (T)}/\nu (T)$ where both $\mu (T)$ and $\nu(T)$ tend to zero
 as $T$ approaches $T_{c}$. We will not discuss anymore this point that was studied in details in \cite{de},\cite{ca}.

\section{Finite $N$ corrections} 

The aim of this section is to compute the free energy taking into account the first corrections in $1/N$.

According to the replica folklore we should be able to compute these corrections in a straightforward way. We should have that  
 \be
 \ov{\ln \left( Z_N (\beta)\right)}\equiv -\beta N F_N(\beta)= -\beta F(Q^*) -\tilde{\mbox{Tr} } (\ln(\beta {\cal H}) ) +\ln(M(m))\ ,
 \ee
 where we have used the short hand notation
 \be
  -\tilde{\mbox{Tr} } (A)= \lim_{n\to 0} {\mbox{Tr }(A(n))\over n} 
  \ee
 and $M(n)$ is a multiplicity factor given by 
 \be
 \lim_{n \to 0} n^{-1} \log(P(n,m))\, ,
 \ee
 where $P(m,n)$ is the number of way in which we can divide $n$ replicas in $m$ groups of $n/m$ replicas.
 As we have seen
 \be
 P(m,n)={n!\over (n/m)! (m!)^{n/m}} \ ,
 \ee
 and therefore
 \be
 M(n)=
 -{\Gamma^{\prime}(1)}
+ \frac{
 \Gamma^{\prime}(1)
- \ln(\Gamma[1+m])} { m}\ .
\ee
 
 The previous formula have a doubtful derivation. In particular it is not clear why one should take the contribution coming from a particular value of $m$ rather than those coming from other values of $m$. We could mumble that the sum over $m$ should become a integral, but is not clear which should be the integration measure.  Moreover we cannot  include  in this way the contribution coming from different values of $m$ because their contribution would be dominant (in the replica approach we have maximized, non minimized the free energy).

Instead the  approach presented in this note  allows us (at least in some cases) to do the computation of the subleading corrections. We will assume that in
 the leading and in the next to leading order the contribution of each block factorizes.  The final 
expression for the contribution of a given saddle point is given by
\be
\exp\left(\sum_{i}^{l} \left( -\beta N m_i f(m_i)-f_{1}(m_{i})\right) \right)  \ .
\ee
where 
\be
f_{1}(m)=\tilde{\mbox{Tr} } (\ln(\beta {\cal H}) ).
\ee

The  $1/N$ corrections arise from two sources.
\begin{itemize}
\item The integral over $y$ has been done approximating the integral with a step function. A more accurate computation, where one consider corrections that are of order (1) for $y-y^*$ of order 1, gives a contribution of order $1/N$ to the free energy density. We call these corrections $\Delta F_{M}$. At this end we must use the
the expression for $C(y)$ shown if eq. (\ref{ephi1}).
\item The corrections coming from the  fluctuations of $Q_{ab} = q_{ab} + \delta q_{ab}$ around the saddle point of solution $Q_{1RSB}$. The final effect of these corrections is denoted $\Delta F_{Q}$ and it equal to $f_{1}(m)$.
\end{itemize}
The computation is similar to that in \cite{de} and we find:

\be
N \beta F_N(\beta) =  N \beta F(x_{sp}) +\Delta F_{M} +
\Delta F_{Q} \ ,
\ee
where $\Delta F_{M}$  is given by
\be
 \Delta F_{M}=(\frac{1}{ m}-1)\Gamma^\prime(1)-\frac{\ln(\Gamma(1-m))}{m}+
  \frac{\ln(2 \pi m^3 \beta N (-f^{\prime \prime}(m)))}{2 m}
 \label{lnz}
\ee

We caution the reader that in many cases the factorization property of the subleading corrections is not true and therefore the value of $\Delta F_Q$ may be incorrect (this point should be carefully investigated). However the distinctive feature of this approach is the presence of the term $\frac{1}{2m} \log(N)$ in the subleading corrections. This is the consequence of having done an extra saddle point integration (with respect to the conventional ones), i.e. the one over $m$: it should be impossible to recover it in the conventional replica approach.  As far as we can see  the presence of such a term (in the free energy) should be a quite general feature of one step replica symmetry breaking and it should be relatively easy to detect it numerically.

In the case of the REM we can verify that we obtain the correct formulae using eq. (\ref{FREM}).
 In this case  terms coming from the determinant are trivial (if we use the replica approach to solve the REM) and therefore we should put  $ \Delta F_{Q} =0$ in the previous equations. Indeed in this case equation (\ref{lnz}) coincide with the expression obtained by
Derrida by an asymptotic expansion of equation (\ref{reprln}).
In this case the equation for $x_{sp}$ can be explicitly solved, obtaining
 $x_{sp} = T/T_c$.
 In more general cases one has a more complicated function $f(m)$ and the saddle point
equations have to be solved numerically. It would be interesting to check if one gets the correct finite volume corrections in simple model like the $p$-spin spherical  model and the $p$-spin Ising model.

\section{Conclusions}

In this note we have shown that by starting from a reasonable form for the leading 
contribution in an exact representation we recover the one step replica broken solution.  We 
introduced a method to calculate the fluctuations over the parameter $m$ of the solution.  In 
doing so,  the parameter $m$ is the saddle point value over an integrating field of 
which we provided the correct measure.  In all our calculations we could check the limit of 
uncorrelated energies (REM) for which we had the solution obtained by Derrida without making use of 
replicas.  Finally, as a side effect, we gave an explanation of the mechanism why in the 1RSB the 
value of $m=T/T_c$ is actually a maximum and not a minimum.

Let us finally remark that the fluctuations on $m$ are important below $T_c$, indeed their 
contribution to the finite-size corrections of the free energy diverges approaching $T_c$ from 
below.  It is possible that these fluctuations  are related to sample-to-sample 
fluctuations of the critical temperature.  In this sense one could say that perturbative corrections 
to $m_{sp}$ reproduce non-perturbative corrections to the matrix  $Q$: when $m$ 
changes, there are some elements $Q_{ab}$ that change abruptly from $q_0$ to $q_1$ and 
vice-versa.  In some works \cite{srpsp} it has already been noted that, in short range models, 
these effects are possibly responsible for the rising of a diverging correlation lengths as 
approaching $T_c$ from above.  It could be interesting to extend our method to short-range models with the aim 
of predicting some of their peculiar features.

\end{document}